\documentclass[onecolumn]{webofc}
\usepackage[varg]{txfonts}   
\usepackage{rotating}
\usepackage{pdflscape}
\usepackage{lscape}
\usepackage{geometry}
\usepackage{lscape}
\usepackage{caption}
\usepackage{titlesec}

\titlespacing*{\section}
{0pt}{5.5ex plus 1ex minus .2ex}{4.3ex plus .2ex}

\begin{document}

\title{The XENON1T Data Distribution and Processing Scheme}

\author{
  	\firstname{Daniel} \lastname{Ahlin}\inst{1}\and 
	\firstname{Boris} \lastname{Bauermeister}\inst{2}\fnsep\thanks{\email{boris.bauermeister@fysik.su.se}} \and
 	\firstname{Jan} \lastname{Conrad}\inst{2}\and 
 	\firstname{Robert} \lastname{Gardner}\inst{3}\and 
 	\firstname{Luca} \lastname{Grandi}\inst{4}\and 
        \firstname{Benedikt} \lastname{Riedel}\inst{3}\and 
        \firstname{Evan} \lastname{Shockley}\inst{4}\and 
        \firstname{Judith} \lastname{Stephen}\inst{3}\and 
        \firstname{Ragnar} \lastname{Sundblad}\inst{1}\and 
 	\firstname{Suchandra} \lastname{Thapa}\inst{3}\and 
 	\firstname{Christopher} \lastname{Tunnell}\inst{4}
}

\institute{
PDC - Center for high performance computing, KTH (Stockholm, Sweden) \and
Oskar Klein Centre, Department of Physics, Stockholm University (Stockholm, Sweden)\and
Enrico Fermi Institute, University of Chicago (Chicago, Illinois) \and
Department of Physics and Kavli Institute for Cosmological Physics, University of Chicago (Chicago, Illinois)
          }

\abstract{
The XENON experiment is looking for non-baryonic particle dark matter in the universe. The
setup is a dual phase time projection chamber (TPC) filled with 3200 kg of ultra-pure liquid xenon. The setup is operated at the Laboratori Nazionali del Gran Sasso (LNGS) in Italy.
We present a full overview of the computing scheme for data distribution and job management in
XENON1T. The software package Rucio, which is developed by the ATLAS collaboration, facilitates data
handling on Open Science Grid (OSG) and European Grid Infrastructure (EGI) storage systems. A tape copy at the Centre for High Performance Computing (PDC) is managed by the Tivoli Storage Manager (TSM). Data reduction and Monte Carlo production are handled by CI
Connect which is integrated into the OSG network. The job submission system connects resources at the EGI,
OSG, SDSC’s Comet, and the campus HPC resources for distributed computing.\\
The previous success in the XENON1T computing scheme is also the starting point for its successor experiment
XENONnT, which starts to take data in autumn 2019.
}

\maketitle

\section{Introduction}
\label{sec-introduction}
Several surveys of our universe show that the amount of baryonic matter does not account for the predicted total amount of matter. Recent studies by the Planck collaboration \cite{planck2015} suggest a matter composition of $\sim$5\% baryonic matter, $\sim$25\% dark matter and $\sim$70\% dark energy. Our understanding of baryonic matter comes from collider experiments, whereas dark matter and dark energy models are broadly discussed in the field of astro-particle physics. Dark matter introduces a new type of non-luminous, massive and electrically neutral particle which is only weakly interacting with baryonic matter. The favoured model is the WIMP (Weakly Interacting Massive Particle) \cite{wimps_dm}.\\
The XENON1T experiment was designed to measure WIMP-nuclei interactions with 3.2 tons of ultra-pure liquid xenon in a two-phase time projection chamber (TPC)\cite{xe1t_2018}. The xenon response to a particle interaction creates a scintillation and ionisation signal which is read out by 248 photomultiplier tubes (PMTs) on the top and bottom of the TPC. The TPC is mounted in a water tank of $\sim$785 $\text{m}^3$ to veto muon introduced events in the TPC \cite{xe1t_expoverview}. These events are tracked by Cherenkov light emission of the muons in the water and the signal is detected by 84 PMTs in the water tank. The experiment is located at the Laboratori Nationali del Gran Sasso (LNGS) in Italy.\\
The XENON1T computing scheme uses several computer centres for data storage and processing. CNAF (Bologna), CCIN2P3 (Lyon), NIKHEF (Amsterdam), SURFsara (Amsterdam), and the Weizmann Institute (Rehovot) are part of the European Grid Infrastructure (EGI) \cite{egi}. SDSC's Comet Supercomputer and the HPC campus resources are connected through the CI Connect infrastructure \cite{ci_connect} to the general Open Science Grid (OSG) \cite{osg}. Independent of grid resources, the Parallelldatorcentrum (PDC, Stockholm) provides tape for long term data storage. The Research Computing Center (RCC) at the University of Chicago hosts the reduced data for interactive analysis and also offers dedicated computing resources.

\section{The XENON1T data overview}
\label{sec-data-review}
\begin{table}[htp]
\centering
\begin{tabular}{|c|c|c|}
\hline
\textbf{Source} & \textbf{Total Amount [TB]} & \textbf{Science run periods [TB]}\\
\hline
Dark Matter & 414.69 & 232.64 \\
LED & 37.37 & 0.0 \\
$^{137}\text{Cs}$ & 8.47 & 0.36\\
$^{83\text{m}}\text{Kr}$ & 62.25 & 29.9\\
$^{220}\text{Rn}$ & 91.14 & 25.64\\
$^{241}\text{AmBe}$ & 68.71 & 62.54\\
$^{228}\text{Th}$ & 3.01 & 0.0\\
Neutron Generator & 54.5 & 10.94\\
MuonVeto & 2.73 & 0.0\\
\hline
\textbf{Total} & 743.23 & 362 \\
\hline
\end{tabular}
\caption{Overview about XENON1T data since commissioning in March 2015. The science run periods were from November 2016 to January 2017 (\textit{SR0}, \cite{xe1t_2017}) and February 2018 to February 2018 (\textit{SR1}, \cite{xe1t_2018}). LED calibrations were regulary performed during science runs to calibrate the light response of the PMTs.}
\label{table:dataoverview}
\end{table}
\noindent Since the XENON1T commissioning in March 2015 the experiment has taken 26762runs. Each run is usually one hour of data taking. The data acquisition system reads and digitises each PMT output and build events from the waveforms. Individual peaks are up to 100 kBin size and one event is up to 2 MBlarge. This output is stored as raw data on disk for further processing. Each run consists of $\sim$20000 events (dark matter data taking mode) and is segmented into sub-files of 100 events.
In addition to the dark matter data, several calibration campaigns are perfomred to study the background in the detector. The total data volume is given in \mbox{table \ref{table:dataoverview}}. XENON1T recorded two long consecutive science data periods (\textit{SR0} \cite{xe1t_2017} and \textit{SR1} \cite{xe1t_2018}). These data volumes are shown separately in \mbox{table \ref{table:dataoverview}}.\\
Several detector related parameters such as pressure or temperature are stored in a \textit{Historian database}. Meta information about the recorded runs, such as run number, number of events, and data locations is stored in a MongoDB database \cite{mongodb}. Both databases are physically hosted at the LNGS. The MongoDB database is mirrored to Stockholm and Chicago to guarantee continuous access for analysts and facilitate raw data processing at connected grid resources.

\section{The XENON1T Software Setup}
\label{sec-software}
The XENON collaboration combines several software approaches for data management, processing and analysis. This section summarises the XENON developments and external software.\\
The data distribution is handled by \textbf{Rucio}, a data management tool which is developed by the ATLAS collaboration \cite{rucio,rucio2}. The XENON collaboration embeds Rucio in its data workflow. The file catalogue, the Rucio server, is hosted at the University of Chicago and the Rucio clients, based on Python 2.7 \cite{python27_software}, are distributed by the CernVM file system (cvmfs)\cite{cvmfs_software} to the connected computer centres.\\
The \textbf{Tivoli Storage Manager (TSM)} manages the tape backup at the PDC in \mbox{Stockholm \cite{tsm_software}}. The TSM server maintenance is done by the PDC and there is one TSM client installed at LNGS. All tape uploads and downloads are put through this single entry point. The tape backup is independent of grid resources.\\
\textbf{PAX} \cite{xenon_collaboration_2018_1195785} is the central processing tool which is developed by the XENON collaboration. The data processor is based on several plugins which are developed for dedicated purposes in the analysis chain. PAX is used at the DAQ for peak identification on the individual waveforms and later classification into scintillation and ionisation signals. PAX runs also at the computer centres to reduce raw data into a smaller data format by extracting physical properties \cite{xe1t_2017}. PAX saves its output in ROOT files \cite{root} using a Python interface to the ROOT I/O format \cite{root_io}.\\
\textbf{HAX} \cite{xenon_hax} is another data processing tool to create further reduced data sets (\textit{minitrees}) for high level analysis. All necessary quantities for data analysis are stored in dedicated minitrees, reducing the overhead, to access analysis level data. Analysts can also create their own minitrees for a particular analysis. HAX saves its output in ROOT files and provides tabular data by using pandas dataframes \cite{software_pandas}.\\
\textbf{CAX} \cite{xenon_cax} is the XENON data management and job submission software and is developed in \mbox{Python 3 \cite{python3x_software}}. It provides several tools to move data by scp, rsync or grid commands, maintenance tools, a job submission system via HTCondor \cite{htcondor} and an interface to Rucio and the TSM client. CAX provides common access to the meta-database. The most important parts for data management and processing in CAX are described below.\\
\textbf{Ruciax} is developed to handle Rucio command line calls by the Python package \textit{subprocess}. This allows interaction between Python 2 and Python 3, which is necessary because Rucio was not compatible with Python 3 at version 1.8.3. Ruciax manages data uploads, downloads and initialises data transfers within the Rucio catalogue (Rucio transfer rules). Ruciax updates the meta-database with latest Rucio transfer rules.\\
\textbf{CAX-Processing} controls the job submission to HTCondor, which submits jobs to OSG (via CI-Connect) or EGI. It uses the meta-database information to determine the status and the latest location of the raw data in the Rucio catalogue to create a DAGMan \cite{htcondor_dag} script for job submission.\\
\pagebreak
\textbf{CAX-TSM} uses the Python package \textit{subprocess} to execute TSM client command line calls. XENON1T raw data are uploaded to tape and registered to the meta-database after a checksum test.\\
Ruciax, CAX-TSM and CAX-Processing are shipped out with CAX. All tools are installed in an Anaconda environment (Python 3.4) and distributed by cvmfs to connected computer facilities. Rucio is installed in another Anaconda environment with Python 2.7.

\section{The XENON1T data flow}
Due to the large size of individual raw data sets and the correspondingly intensive computational effort, the XENON collaboration decided to use a data flow model with three stages: \textit{raw data}, \textit{processed data} and \textit{minitrees}. Each data product is created from its previous stage and data are reduced step-wise to get smaller products which can be easily accessed by the analysts.\\ 
The first stage (raw data) consists of the digitised waveform information, which is temporally stored at the DAQ stage. The raw data are the basis for further processing and are kept on dedicated grid storage locations for fast access. Quantities, such as energy and event positions are determined at the second stage. The computationally more expensive calculations at this stage are calculated by PAX at computer centres. Since XENON1T keeps the full raw data information (waveforms) there is only a single copy of the processed data necessary, which is stored at RCC. At the last step, the processed data are reduced to minitrees and further quantities such as corrected ionisation and scintillation signals are calculated. The minitree creation is less computationally intensive and is executed at RCC. In case of changing initial corrections or newly added quantities, minitrees can be regrown at any time at RCC. Minitrees are the final data product and are distributed to the data analysts. A single raw data set is described by several (predefined) minitrees and are sorted by categories such as basic information, interaction types, positions, and corrections. Each individual minitree is much smaller in data size which allows analysts to work independently of computer centres.
\label{sec-data-workflow}
\begin{figure}[ht]
    \centering
    \includegraphics[width=1\textwidth, angle=0, scale=0.95]{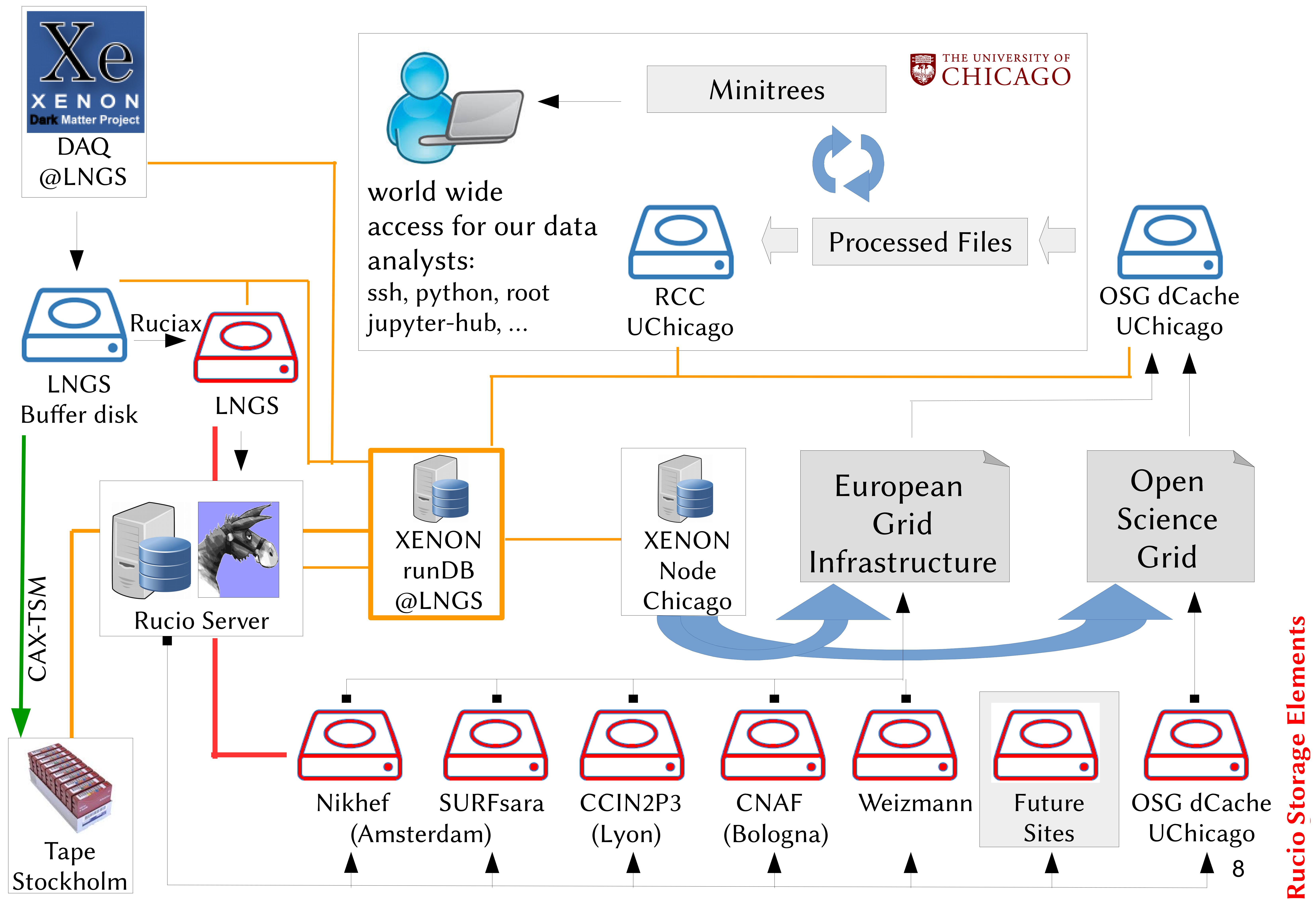}
    \caption{Overview of the XENON1T data flow. Incoming raw data are uploaded at the LNGS to the Rucio catalogue (Ruciax) and distributed according to predefined data transfer rules (red connectors). The first data reduction is done on EGI or OSG with PAX. Another processing step (HAX) is then executed at RCC. The final product, the minitrees, are available at RCC for XENON1T analysts. Locations of raw and processed data are tracked in the XENON1T meta-database (yellow connectors). Minitrees are not tracked by the meta-database. An independent tape backup stores data at the PDC in Stockholm (green connector).}
    \label{the_dataflow}
\end{figure}

\subsection{The XENON1T data distribution scheme}
\label{sec-data-distribution}
The complete data distribution scheme is shown in figure \ref{the_dataflow}. The data distribution begins with the raw data product which is written by the DAQ to a 50 TB buffer disk at the LNGS (LNGS buffer disk). The buffer disk allows continuously data taking for several days in case of network interruption to the external data storage.\\
Ruciax handles the upload of raw data to the Rucio catalogue. The raw data are written to another buffer disk at the LNGS, which acts as a \textit{Rucio Storage Element} (RSE) within Rucio. The raw data lifetime at this RSE is usually four days. Ruciax initiates the data distribution to further worldwide distributed RSEs for data storage. Four days are enough to fulfil data transfers to other RSEs. The meta-database is updated with the RSE information by Ruciax when a data transfer is fulfilled.\\
Once a raw data set is registered to another RSE in the meta-database, the data are ready for processing with OSG (daily processing). After the processed files are collected at OSG dCache in Chicago, they are transferred to RCC. Processed data are transferred outside the Rucio catalogue with CAX and the underlying \textit{scp} command. The processed data are mainly treated as an intermediate product which is usually not used by the analysts. Therefore, they are not distributed with the Rucio catalogue. At the last step, the meta-database is updated with the latest locations of the processed data at RCC.\\
The last processing stage creates the minitrees at RCC, which are not tracked in the meta-database. They are moved within RCC to dedicated folders, which are accessible to analysts.\\
The tape storage at the PDC is not connected to any grid resources and is treated as an independent storage location. The CAX-TSM instance takes care of uploads and downloads. Raw data are transferred to a tape storage location as soon as they arrive at the LNGS buffer disk. That happens in parallel to the Rucio uploads. In case of an unforeseen data loss in the Rucio catalogue, we can restore raw data to the LNGS buffer disk and re-upload them to the Rucio catalogue again.

\subsection{The XENON1T processing scheme}
\label{sec-data-processing}
\begin{figure}[tbp]
    \centering
    \includegraphics[width=0.99\textwidth, scale=1]{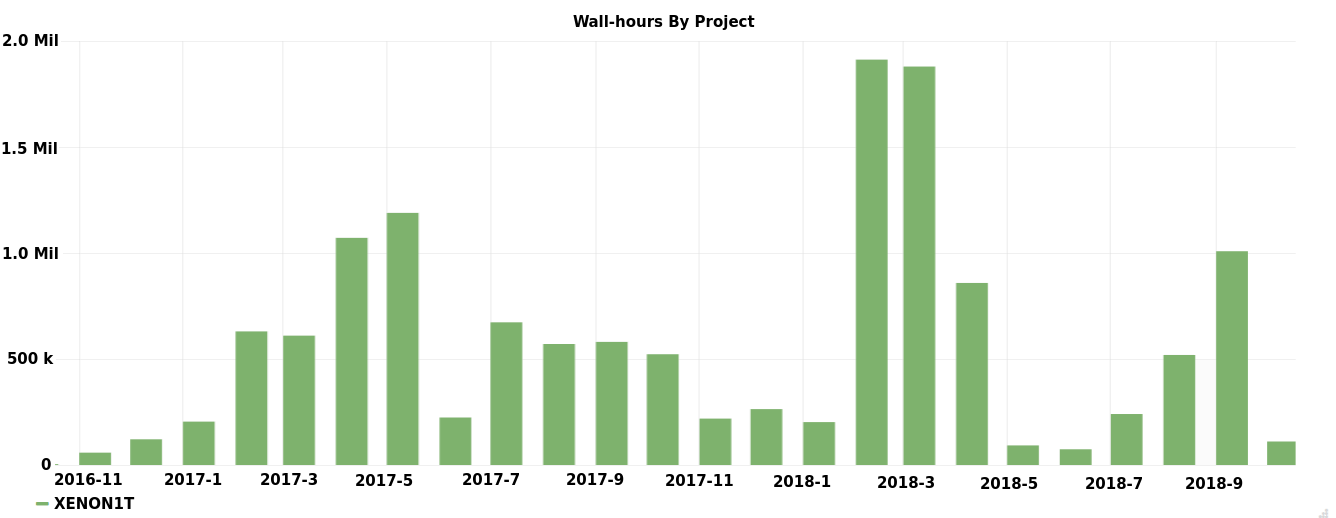}
    \caption{XENON1T wall-hours overview from all connected computing resources. Peaks such as May 2017, March 2017 and September 2018 originate from  reprocessing campaigns. Further usage is daily processing and Monte Carlo simulations of the XENON1T TPC and the future XENONnT TPC.}
    \label{the_wallhours}
\end{figure}
Processing raw data in XENON1T consists of \textit{daily processing} and \textit{scheduled reprocessing campaigns}. In both cases, the job submission software (CAX) sends processing jobs via HTCondor. All jobs are submitted from the XENON node in Chicago. The decision at which computer centre the data are processed depends on the location in the Rucio catalogue and is set in the DAGMan scripts. OSG is preferred over EGI for daily processing tasks since dCache Chicago holds a copy of the latest raw data.\\
To increase the throughput and distribute the job processing more equally, a single raw data set is not processed on a single cluster machine. By default, the raw data are separated into chunks of 100 events each. The job submission handler splits the job by creating \textit{spliced DAGs} within one DAGMan script. This requires only one job submission per data set, which processes all chunks of data in parallel. Once all individual data chunks are processed, they are merged into the final processed file. This file is temporally stored at dCache Chicago and transferred to RCC. Failing jobs are submitted again automatically.\\
Major updates in PAX require larger reprocessing campaigns which are performed in addition to the daily processing. Therefore, we use the resources of all connected EGI and OSG sites depending on the location of a raw data set in the Rucio catalogue. Figure \ref{the_wallhours} shows the wall-hours usage of all connected computing resources.\footnote{Retrieved from https://gracc.opensciencegrid.org/dashboard/db/payload-jobs-summary} Peaks such as May 2017, March 2018 or September 2018 are reprocessing campaigns. Further large workload comes from Monte Carlo simulations.

\subsection{Data handling policy}
\label{sec-data-handling-policy}
To assure high data safety, the XENON collaboration uses mainly two data storage policies to prevent data loss. Since data can be reprocessed at any time, the data loss policy applies only to raw data. Data is categorised into \textit{science run} and \textit{non-science run} data. The latter is raw data which were taken during detector commissioning phase or data which are in general not used for dark matter analysis.\\
Science run data require a three-fold data safety: at dCache Chicago RSE, a random RSE in Europe and a valid tape copy. Non-science run data require only a two-fold copy: a random RSE in Europe and a valid tape copy.\\
All data safety mechanisms are based on an incorruptible upload of the initial raw data set to the LNGS buffer disk (RSE). Therefore, data are only purged with a dedicated tool from the LNGS buffer disk. This tool always requests a valid tape copy with a checksum test and asks for at least one valid copy in Rucio which is not located at the LNGS RSE.

\section{Summary}
\label{sec-data-summary}
The XENON collaboration opts for a mainly Python-based software solution to implement the data management and processing workflow. The computing concept has been well-established since the commissioning phase of XENON1T and has proven to be a reliable system. We have introduced a high data safety policy to prevent data loss with a two and three-fold backup.\\
The existing computing concept is under revision due to upcoming changes for XENONnT. The XENON collaboration continues with the current grid-based data distribution in Rucio and the job submission system under the usage of a meta-database.

\end{document}